\documentclass[a4paper,12pt]{article} 

\usepackage[english]{babel} 

\usepackage{amssymb}
\usepackage{setspace}
\usepackage{afterpage}
\usepackage{latexsym}
\usepackage{amsmath}
\usepackage{amsthm}
\usepackage{amsfonts}
\usepackage{mathrsfs}
\usepackage{graphicx}
\usepackage{hyperref}

\newcommand{\lR}{\mathrm{I\hspace{-0.7mm}R}}

\numberwithin{equation}{section}

\hypersetup{ pdftitle={Draft},
pdfauthor={Bobby Eka Gunara and Ainol Yaqin}, 
pdfkeywords={Nonminimal Derivative Coupling Gravity}, 
bookmarksnumbered, pdfstartview={FitH}, urlcolor=blue, }

\begin{document}
\pagestyle{plain}


 \title{\LARGE \textbf{Self-Gravitating Static Non-Critical  Black Holes in 4$D$ Einstein-Klein-Gordon System with Nonminimal Derivative Coupling}}

\author{{ Bobby Eka Gunara$^{\flat,\sharp}\footnote{Corresponding author}$ and  Ainol Yaqin$^{\sharp}$ } \\
$^{\flat}$\textit{\small Indonesian Center for Theoretical and
Mathematical Physics (ICTMP)}
\\ {\small and} \\
$^{\sharp}$\textit{\small Theoretical Physics Laboratory}\\
\textit{\small Theoretical High Energy Physics and Instrumentation Research Group,}\\
\textit{\small Faculty of Mathematics and Natural Sciences,}\\
\textit{\small Institut Teknologi Bandung}\\
\textit{\small Jl. Ganesha no. 10 Bandung, Indonesia, 40132}\\
\small email: yaqin.al@students.itb.ac.id, bobby@fi.itb.ac.id\\
}

\date{}

\maketitle

\begin{abstract}
 We study static non-critical hairy black holes  of four dimensional gravitational model with nonminimal derivative coupling and a scalar potential turned on. By taking an ansatz, namely,  the first derivative of the scalar field  is proportional to  square root of  a metric function, we reduce  the Einstein field equation  and the scalar field equation of motions  into a single highly nonlinear differential equation. This setup implies that the hair is secondary-like since the scalar charge-like depends on the non-constant mass-like quantity in the asymptotic limit. Then, we show that near boundaries the solution is not the critical point of the scalar potential and the effective geometries become spaces of constant scalar curvature. 
 \end{abstract}

\section{Introduction}
\label{sec:intro}

Modified Einstein's gravitational theories  have been  intensely studied over the last thirty years since they might give a solution to some cosmological problems such as the dark matter and dark energy, the inflationary scenario in the early universe, and the accelerated expansion of our  universe.  Among the models, it is of interest to consider a model  called non minimal derivative coupling (NMDC) gravitational theory in four dimensions since the model has been shown to provide an inflationary accelerated expansion universe without introducing any scalar potential \cite{Amendola:1993uh}. The action of the theory has the form \cite{Amendola:1993uh, Horndeski:1974wa}
\begin{equation}  \label{ActionNMDC} 
S = \frac{1}{2} \int d^{4} x\sqrt{- g } \left[\frac{R-2\Lambda }{\kappa ^{2} } -\left( \varepsilon  g_{\mu \nu } + \xi  R g_{\mu \nu } + \eta  R_{\mu \nu } \right)\partial ^{\mu } \phi \partial ^{\nu } \phi  \right]  
\end{equation} 
where $\phi$ is a real scalar field, $g_{\mu \nu }$ is the spacetime metric, and $g \equiv \mathrm{det}(g_{\mu \nu })$, while $R_{\mu \nu }$    and $R$ are  the Ricci tensor and the scalar curvature, respectively. In the rest of this paper we use the indices $\alpha, \beta, \mu, \nu = 0,...3$.  The constants  $\xi$ and $\eta$ are real and called the derivative coupling parameter with the dimension of length-squared. Whereas $\Lambda$ is the cosmological constant and $\kappa \equiv 1/m_p$ with $m_p$ is the Planck mass. The parameter $\varepsilon = \pm 1$, where $\varepsilon = 1$ correspond to a theory with a canonical scalar field, while  $\varepsilon = - 1$ correspond to a theory with a phantom scalar field \cite{Korolev:2014hwa}.

The action in (\ref{ActionNMDC}) admits a global translational symmetry for the scalar field $\phi \to \phi + a$ where $a$ is an arbitrary real constant on the four dimensional curved spacetime. In the case of static spacetimes there have  been several studies revealing that black hole solutions in the theory may have hair, see a review for example in  \cite{Babichev:2016rlq, Volkov:2016ehx}. Most of the hair are secondary because the scalar field depends explicitly on the mass of the black hole while the mass together with its scalar charge are fixed. However, such a situation might differ when we replace the cosmological constant $\Lambda$ in (\ref{ActionNMDC})  by a scalar potential $V(\phi)$. For example, this scalar potential when the theory is nonminimally coupled  to gravity, could be viewed as the source of dark matter decays which has been recently considered in \cite{Cata:2016dsg}. Moreover, the dark matter lifetime could be at least of the same order as the age of the Universe, which is quite interesting in the context of the standard cosmological model.\\
%
%
%
%
\indent Apart from the cosmological context mentioned above, it is also of interest to consider  a class of static black hole solutions in the four dimensional Einstein-scalar theory with nonminimal derivative coupling and the scalar potential $V(\phi)$ turned on. The spacetime metric of the theory is conformal to  ${\bf M}^2 \times {\bf S}^2_k$ where ${\bf M}^2$ is a two-surface, while  ${\bf S}^2_k$ are Einstein surfaces with $k = 0, \pm 1$.  The latter cases are related to two-torus, two-sphere, and Riemann surfaces as will be discussed in the next section. To solve the Einstein field equation and the scalar equation of motions, we take an ansatz on the first derivative of the scalar field such that it is proportional to the square root of the metric function, see  (\ref{simpelphi}). Such a setup  simplifies the Einstein field equation  and the scalar field equation of motions  into a single highly nonlinear differential equation in $Y$ where $Y$ is a metric function. Although, it is difficult to solve this equation, it is of interest to focus on the behavior of the solutions near the boundaries, namely in the asymptotic region and near the horizon limit.\\
\indent  In the asymptotic limit, we may have that at the zeroth order the function $Y$ is fixed, but the scalar $\phi$  cannot be frozen. This implies that  the scalar potential cannot be extremized. Two quantities called  mass-like and scalar charge-like of  black holes are not constant and depend on the radial coordinate.  This situation leads us to introduce the notion of \textit{secondary-like} hair for the scalar $\phi$.  The ``effective'' geometry turns out to be a space of constant scalar curvature but not maximally symmetric. This space could be Einstein with negative cosmological constant or a space of negative constant scalar curvature (see Section \ref{sec:solbond}).\\ 
\indent Near-horizon limit the surfaces   ${\bf S}^2_k$  become minimal or  $Y = \epsilon$ with $0 < \epsilon \ll 1 $. This would lead to the blow up of the ansatz in (\ref{simpelphi}) unless if the first derivative of the metric function $\rho(r)$ with respect to the radial coordinate $r$ is proportional to $\epsilon^{\frac{1}{2}}$ where $\rho(r)$ can be thought of as the ``radius'' of  ${\bf S}^2_k$. The near-horizon geometry in this case is $\lR^{1,1} \times {\bf S}^2_k$ where $\lR^{1,1}$ is the 2-Minkowskian surface for $k = \pm 1$. The scalar field $\phi$ in this case depends linearly on $r$ which implies that it is not the critical point of the scalar potential. It turns out that the surface gravity vanishes in this case which follows that we have an ultra-cold black hole \cite{Kallosh:1992}.  This situation differs from the minimal derivative coupling model considered, for example in \cite{ Gibbons:1996prl, Gunara:2010iu} and the non minimal derivative coupling model  discussed  in \cite{Rinaldi:2012vy}.\\
%
%
%
%
%
\indent The structure of the paper can be mentioned as follows. In section \ref{sec:NMDC} we give a short review on gravitational theory with NMDC and the scalar potential.  We also shortly discuss the Einstein field equation and the scalar equation of motions on static  spacetimes. In section \ref{sec:specsol} we consider a special exact solution by setting the ansatz mentioned above and then, derive a highly nonlinear differential equation called the master equation. Then, we discuss the behavior of the master equation near the boundaries in section \ref{sec:solbond}. Finally, we conclude our results in Section \ref{sec:conc}.

\section{A Short Review on NMDC Gravitational Theory}
\label{sec:NMDC}
\subsection{General Setup}
In this subsection, we give a quick review on four dimensional NMDC Einstein-Klein-Gordon theory  with scalar potential turned on. The theory  is described by an action of the form 
\begin{equation} \label{ActionNMDCV} 
S = \frac{1}{2} \int d^{4} x  \sqrt{-{\it g}}  \left[\frac{1}{\kappa ^{2} } R - \left( \varepsilon g_{\mu \nu } + \eta G_{\mu \nu } \right) \partial ^{\mu } \phi  \partial ^{\nu } \phi - 2 V(\phi ) \right] \ ,
\end{equation} 
where  $G_{\mu \nu }$ is the Einstein tensor whose form is given by
\begin{equation}\label{Einsteineq}
G_{\mu \nu } \equiv R_{\mu \nu } - \frac{1}{2}  g_{\mu \nu } R  ~ ,
\end{equation}
and $V(\phi )$ is the real scalar potential. In this case, the pre-coefficients  $\eta$ and $\xi$ have been chosen such that $-2 \xi = \eta$.

Varying (\ref{ActionNMDCV}) with respect to the metric $g_{\mu \nu }$, we get the Einstein field equation
\begin{equation}
G_{\mu \nu } = \kappa ^{2} \left( T_{\mu \nu }^{(\phi )} +T_{\mu \nu }^{(\eta)} \right)  ~ ,
\end{equation}
with
\begin{eqnarray}
T_{\mu \nu }^{(\phi )}  &=& \varepsilon \nabla_{\mu } \phi \nabla_{\nu } \phi - \frac{\varepsilon }{2} g_{\mu \nu } \nabla^{\alpha } \phi  \nabla _{\alpha } \phi - g_{\mu \nu } V(\phi)  ~  , \nonumber\\
 \frac{1}{\eta} T_{\mu \nu }^{(\eta )} &=& R_{\mu \alpha } \nabla_{\nu } \phi \nabla^{\alpha } \phi + R_{\alpha \nu } \nabla^{\alpha } \phi \nabla_{\mu } \phi + \frac{1}{2} g_{\mu \nu } \nabla^{\alpha } \nabla^{\beta } \phi \nabla_{\alpha } \nabla _{\beta } \phi + \frac{1}{2} g_{\mu \nu } \left(\nabla _{\alpha } \nabla ^{\alpha } \phi \right)^2 \nonumber\\
 &&  -  \frac{1}{2} G_{\mu \nu } \nabla^{\alpha } \phi  \nabla _{\alpha } \phi  
-\nabla_{\mu } \nabla_{\nu } \phi \nabla _{\alpha } \nabla ^{\alpha } \phi + R_{\mu \alpha \nu \beta } \nabla^{\alpha } \phi \nabla^{\beta } \phi +\nabla_{\mu } \nabla^{\alpha } \phi \nabla_{\nu } \nabla _{\alpha } \phi \nonumber\\
&& - \frac{1}{2} R \nabla _{\mu } \phi  \nabla _{\nu } \phi
  - g_{\mu \nu } \nabla^{\alpha } \nabla^{\beta } \phi \nabla _{\alpha } \nabla_{\beta } \phi - g_{\mu \nu } R^{\alpha \beta } \nabla _{\alpha } \phi \nabla_{\beta } \phi  ~  .
 \end{eqnarray}
The scalar field equation of motions can be obtained by varying (\ref{ActionNMDCV}) with respect to $\phi$, namely
\begin{equation}\label{scalareqom}
 \left( \varepsilon g^{\mu \nu } + \eta G^{\mu \nu } \right)\nabla _{\mu } \nabla _{\nu } \phi =\frac{dV(\phi )}{d\phi } ~ ,
\end{equation}
where we have used $\nabla_{\mu} G^{\mu \nu } = 0$. 

\subsection{Static  Spacetimes}
In this subsection we focus  on a case of  static spacetimes whose metric has the  form
\begin{equation} \label{staticmet} 
ds^2 = - f(r) dt^2 + g(r) dr^2 + \rho^{2} (r) d\Omega^{2}_k  ~  ,
\end{equation} 
where $r$ is the radial coordinate and $d\Omega^2_k $ describes two dimensional Einstein surfaces with $k = 0, \pm 1$. For $k=-1$, $k=0$, and $k=1$, we have Riemann surface, the two-torus, and the two-sphere, respectively. Note that among the functions $f(r)$, $g(r)$, and $\rho(r) $, only two of them are independent. This can be seen by employing a redefinition of the radial coordinate $r$ to absorb one of them.  It is important to write down the norm of the Riemann tensor related to the metric (\ref{staticmet}), namely
\begin{equation} \label{normRiem} 
R^{\alpha\beta\mu\nu} R_{\alpha\beta\mu\nu} =  \frac{3}{2 f^2 g^2} \left( f'' - \frac{f'^2}{2f}-\frac{f' g'}{2g}\right)^2 + \frac{3 \rho'^2 f'^2}{\rho^2 f^2 g^2} + \frac{12}{\rho^2  g^2} \left( \rho'' -\frac{\rho' g'}{2g}\right)^2 + \frac{6}{\rho^4}  \left( k -\frac{\rho'^2}{g}\right)^2  ~  ,
\end{equation} 
which will be useful for our analysis in the next section. Moreover, throughout this paper we simply assume that the scalar field $\phi$ in (\ref{Einsteineq}) - (\ref{scalareqom}) depends only on $r$, i.e. $\phi = \phi(r)$.

The above setup simplifies  the field equations (\ref{Einsteineq}), to become
\begin{eqnarray} 
\frac{\rho ' g'}{\rho f g^2 } -\frac{\rho '^2 }{\rho^2 fg} -\frac{2 \rho ''}{\rho fg} + \frac{k}{\rho^2 f} &=& \eta \kappa ^{2} \left(\frac{3\rho 'g'}{2\rho fg^{3} } -\frac{\rho '^{2} }{2\rho ^{2} fg^{2} } -\frac{k}{2\rho ^{2} fg} -\frac{\rho ''}{\rho fg^{2} } \right)\phi '^{2}  \nonumber \\ 
&& + \frac{\varepsilon \kappa^2 }{2} \frac{1}{fg} \phi '^2  - \eta \kappa ^2 \frac{2\rho '}{\rho fg^{2} } \phi '\phi '' + \kappa ^{2} \frac{V}{f}  ~  ,\nonumber \\ 
\frac{\rho 'f'}{\rho f} +\frac{\rho '^{2} }{\rho ^{2} } -\frac{k g}{\rho ^{2} }  &=& \frac{1}{2} \kappa ^{2} \varepsilon \phi '^{2}  -\kappa ^{2} gV+\eta \kappa ^{2} \left(\frac{3\rho '^{2} }{2\rho ^{2} g} +\frac{3\rho 'f'}{2\rho fg} -\frac{k}{2\rho ^{2} } \right)\phi '^{2} \nonumber 
 \end{eqnarray}
\begin{eqnarray} 
\frac{f''}{2f^2 g} -\frac{f'g'}{4f^{2} g^{2} } -\frac{f'^{2} }{4f^{3} g} +\frac{\rho 'f'}{2\rho f^{2} g} -\frac{\rho 'g'}{2\rho fg^{2} } +\frac{\rho ''}{\rho fg}  & =& -\frac{\varepsilon \kappa ^{2} }{2} \frac{1}{fg} \phi '^{2}  -\kappa ^{2} \frac{V}{f} \nonumber\\
 && +\eta \kappa ^{2} \left(\frac{f'}{2f^{2} g^{2} } +\frac{\rho '}{\rho fg^{2} } \right)\phi '\phi '' \nonumber\\
 && +\eta \kappa ^{2} \Bigg(\frac{f''}{4f^{2} g^{2} } -\frac{f'^{2} }{8f^{3} g^{2} } -\frac{3f'g'}{8f^{2} g^{3} }  \nonumber\\
  && + \frac{\rho ''}{2\rho fg^{2} } +\frac{\rho 'f'}{4\rho f^{2} g^{2} } -\frac{3\rho 'g'}{4\rho fg^{3} } \Bigg)\phi '^{2} \nonumber\\ 
  \label{Einsteineqstatic}
 \end{eqnarray}
and the scalar field equation of motions (\ref{scalareqom}), to become 
\begin{equation}\label{scalareqomstatic}
\frac{1}{\rho ^{2} \sqrt{fg} } \left\{\frac{\sqrt{fg} }{g} \left[\varepsilon \rho ^{2} +\eta \left(\frac{\rho \rho 'f'}{fg} +\frac{\rho '^{2} }{g} - k \right)\right]\phi '\right\}' = \frac{dV(\phi )}{d\phi }  ~  ,
\end{equation}
where we have defined $A' \equiv dA/dr$ and $A'' \equiv d^2A/dr^2$ for any $A \equiv A(r)$.



\section{A Special Class of  Solutions}
\label{sec:specsol}

\noindent In this section we discuss a simple model where we choose the following ansatz for the scalar field 
\begin{equation} \label{simpelphi} 
\phi' = \nu g^{1/2}  ~  ,
\end{equation} 
where $\nu$ is a non zero real constant and its value is determined by the field equations in (\ref{Einsteineqstatic})  and (\ref{scalareqomstatic}).  The ansatz  (\ref{simpelphi})  is inspired by the scalar-torsion theory \cite{Kofinas:2015hla} which can be viewed as the solutions of its field equations of motions. As will be seen in the next section, this setup has a  non-critical hairy black hole solution whose hair is said to be \textit{secondary-like}, since the ansatz (\ref{simpelphi}) does not yield any critical solution near the boundaries and moreover, the  mass-like quantity of the black hole cannot be a constant but rather depends on $\rho$ in the asymptotic region.  Thus this case is different than the case of critical hairy black holes, see for example  \cite{Gibbons:1996prl, Gunara:2010iu}.

Let us first discuss some consequences of  (\ref{simpelphi}).  The condition  (\ref{simpelphi}) simply casts  (\ref{Einsteineqstatic}) and \eqref{scalareqomstatic} into
\begin{eqnarray} 
\frac{\rho '}{\rho } \left(\frac{g'}{g} -\frac{\rho '}{\rho } \right)-\frac{k g}{\rho ^{2} } -\frac{2\rho ''}{\rho } &=& \frac{\kappa ^{2} }{2-\eta \kappa ^{2} \nu ^{2} } \left(\varepsilon \nu ^{2}  + 2V-\frac{4k}{\kappa ^{2} \rho ^{2} } \right)g ~  , \nonumber\\
\frac{2-3\eta \nu ^{2} \kappa ^{2} }{\kappa ^{2} } \frac{\rho '}{\rho } \left(\frac{f'}{f} +\frac{\rho '}{\rho } \right) &=& -\left(-\varepsilon \nu ^{2}  + 2V - \frac{ (2-\eta \nu ^{2} \kappa ^{2}) k }{\kappa ^{2} \rho ^{2}}  \right)g  ~ , \nonumber\\
\frac{\rho '}{2\rho } \left(\frac{g'}{g} -\frac{f'}{f} \right)+\frac{f'}{4f} \left(\frac{g'}{g} +\frac{f'}{f} \right)-\frac{f''}{2f} -\frac{\rho ''}{\rho } &=& \frac{\kappa ^{2} }{2-\eta \kappa ^{2} \nu ^{2} } \left(\varepsilon \nu ^{2}  + 2V\right) ~ , \nonumber\\
\left\{\nu \sqrt{f} \left[\varepsilon \rho ^{2} +\eta \left(\frac{\rho \rho 'f'}{fg} +\frac{\rho '^{2} }{g} - k \right)\right]\right\}^{{'} }  &=& \rho ^{2} \sqrt{fg} \frac{dV(\phi )}{d\phi }  ~ . \label{simplyEOM}
\end{eqnarray} 
Then, by employing the transformation $f'=\nu \sqrt{g} \dot{f}$ and  $f''=\nu ^{2} \left(\ddot{f}g+\frac{\dot{f}\dot{g}}{2} \right)$ where $\dot{f} \equiv \frac{df}{d\phi } $, $\ddot{f} \equiv \frac{d^{2} f}{d\phi ^{2} } $, we can rewrite   \eqref{simplyEOM} as
\begin{eqnarray} 
 \frac{\ddot{\rho }}{\rho } +\frac{\dot{\rho }^{2} }{2\rho ^{2} } +\frac{k}{2\nu ^{2} \rho ^{2} } &=& -\frac{\kappa ^{2} }{2\nu ^{2} \left(2-\eta \kappa ^{2} \nu ^{2} \right)} \left(\varepsilon \nu ^{2} + 2V-\frac{4 k}{\kappa ^{2} \rho ^{2} } \right) ~ , \nonumber\\
 \frac{\dot{\rho }}{\rho } \left(\frac{\dot{f}}{f} +\frac{\dot{\rho }}{\rho } \right) &=& - \frac{\kappa ^{2} }{\nu ^{2} \left(2-3\eta \nu ^{2} \kappa ^{2} \right)} \left(-\varepsilon \nu ^{2} + 2V - \frac{ (2-\eta \nu ^{2} \kappa ^{2} ) k}{\kappa ^{2} \rho ^{2}}  \right)  ~  ,  \nonumber\\
 \frac{\ddot{\rho }}{\rho } +\frac{\dot{f}}{f} \left(\frac{\dot{\rho }}{2\rho } -\frac{\dot{f}}{4f} \right)+\frac{\ddot{f}}{2f}  &=& \frac{\kappa ^{2} }{\nu ^{2} \left(2-\eta \kappa ^{2} \nu ^{2} \right)} \left(\varepsilon \nu ^{2} + 2V\right)  ~  ,  \nonumber\\
 \left\{\sqrt{f} \rho ^{2} \left[\varepsilon +\eta \nu ^{2} \frac{\dot{\rho }}{\rho } \left(\frac{\dot{f}}{f} +\frac{\dot{\rho }}{\rho } \right)-\eta \frac{k}{\rho ^{2} } \right]\right\}^{.}  &=& \frac{\rho ^{2} \sqrt{f} }{\nu ^{2} } \dot{V}   ~  .  \label{simplyEOM1}
\end{eqnarray} 
Now, in order to get the exact solutions we introduce new variables following \cite{Kofinas:2015hla}. First, we define 
\begin{eqnarray} 
 x &=& \ln \rho ~  ,  \nonumber\\
 y &=& \frac{\dot{\rho }}{\rho } ~  ,  \nonumber\\
 z &=& \frac{\left(\rho f\right)^{.} }{\rho f}  ~ ,
 \end{eqnarray} 
 which implies that  the set of equation in  \eqref{simplyEOM1} becomes
\begin{eqnarray} 
&& \dot{y}+\frac{3}{2} y^{2} +\frac{k}{2\nu ^{2} } e^{-2x} +\frac{\kappa ^{2} }{2\nu ^{2} \left(2-\eta \kappa ^{2} \nu ^{2} \right)} \left(\varepsilon \nu ^{2} + 2V-\frac{4 k}{\kappa ^{2} } e^{-2x} \right) = 0  ~ , \nonumber\\
zy &=& \frac{\kappa ^{2} }{\nu ^{2} \left(2-3\eta \nu ^{2} \kappa ^{2} \right)} \left(\varepsilon \nu ^{2}  - 2V + (2-\eta \nu ^{2} \kappa ^{2}) \frac{  k }{\kappa ^{2} } e^{-2x} \right)  \equiv F ~ , \nonumber\\
&& \dot{z}+\frac{1}{2} z^{2} -\frac{k}{2\nu ^{2} } e^{-2x} +\frac{\kappa ^{2} }{2\nu ^{2} \left(2-\eta \kappa ^{2} \nu ^{2} \right)} \left(3\varepsilon \nu ^{2} +6V+\frac{4k}{\kappa ^{2} } e^{-2x} \right) = 0      ~ , \nonumber\\
&& \left(yz\right)^{.} -\frac{1}{\eta \nu ^{4} } \dot{V}+\frac{k}{2\nu ^{2} } \left(y-z\right)e^{-2x} +\frac{1}{2} \left(3y+z\right)\left(\frac{\varepsilon }{\eta \nu ^{2} } +yz\right) = 0  ~ .  \label{simplyEOM2}
\end{eqnarray} 
In this form, it is easy to see that the second equation in (\ref{simplyEOM2}) can be viewed as a constraint, while the last equation in (\ref{simplyEOM2})  is superfluous  \cite{Kofinas:2015hla}.

Second, we introduce again a set of new variables
\begin{eqnarray}
 Y &\equiv& y^2  ~ , \nonumber \\ 
 Z &\equiv& z^2  ~ ,
 \end{eqnarray}
such that we could have \cite{Yaqin:2017bij} 
\begin{eqnarray}
\frac{1}{2} \frac{dY}{dx}  &=& \dot{y} ~ , \nonumber \\ 
 \frac{1}{2} \varsigma \sqrt{ \frac{Y}{Z}} \frac{dZ}{dx}   &=&  \dot{z}  ~ , \label{thecorrectone}
 \end{eqnarray}
where $\varsigma = \pm 1$.  Then, (\ref{simplyEOM2}) can be rewritten as \cite{Kofinas:2015hla, Yaqin:2017bij}
\begin{eqnarray}
\frac{dY}{dx} +3Y+\frac{k}{\nu^{2} } e^{-2x} + \frac{\kappa ^{2} }{\nu ^{2} \left(2-\eta \kappa ^{2} \nu ^{2} \right)} \left(\varepsilon \nu ^{2}  + 2V-\frac{4 k}{\kappa ^{2} } e^{-2x} \right) &=& 0 ~ , \nonumber \\ 
\frac{\kappa^4 }{\nu^4  \left(2 - 3 \eta \kappa ^{2} \nu ^{2} \right)^2 }  \left( \varepsilon \nu ^{2}  - 2 V +\frac{k}{\kappa ^{2} } (2 - \eta \kappa ^{2} \nu ^{2}  ) e^{-2x} \right)^2  &=& YZ ~ , \nonumber \\ 
\varsigma \sqrt{YZ} ~ \frac{dZ}{dx} + Z^2 +\frac{\kappa ^{2} }{\nu ^{2} \left(2-\eta \kappa ^{2} \nu ^{2} \right)} \left(3\varepsilon \nu ^{2}  + 6V +\frac{k}{\kappa ^{2} } (2 + \eta \kappa ^{2} \nu ^{2}  ) e^{-2x} \right)Z &=& 0  ~ . \label{simplyEOM3}
\end{eqnarray}
It is important to mention that there is a crucial miscalculation in transforming the equation of motions (\ref{simplyEOM1}) into the new variables $(Y, Z)$ in ref. \cite{Kofinas:2015hla} which has been corrected in \cite{Yaqin:2017bij}. The source of the error is the missing prefactor $\varsigma \sqrt{ \frac{Y}{Z}}$ in the second equation of (\ref{thecorrectone}). \\
\indent After some computations using the first and the second equations in (\ref{simplyEOM3}) we get the following results
\begin{eqnarray}
Z &=& \frac{\tilde{\chi }^{2} }{\chi ^{2} Y} \left( \frac{dY}{dx} + 3Y - 2 k \eta \chi \nu^2 e^{-2x} +2\varepsilon \chi \nu ^{2} \right)^{2} ~  ,\nonumber\\
V &=& -\frac{1}{2\chi } \left( \frac{dY}{dx} +3Y\right)-\frac{\varepsilon \nu ^{2} }{2}  -\frac{k}{2 \kappa ^{2} } (2 + \eta \kappa ^{2} \nu ^{2}  )  e^{-2x}  ~ ,  \label{simplyEOM4}
\end{eqnarray}
where $\chi$ and $\tilde{\chi }$  are constants defined as
\begin{eqnarray}
\chi &\equiv&  \frac{\kappa ^{2} }{\nu ^{2} \left(2-\eta \nu ^{2} \kappa ^{2}  \right)}   ~  ,\nonumber\\
\tilde{\chi }  &\equiv& \frac{\kappa ^{2} }{\nu ^{2} \left(2-3 \eta \nu ^{2} \kappa ^{2} \right)} ~ ,
\end{eqnarray}
which are non-singular, \textit{i.e.} $\eta \kappa ^{2} \nu ^{2} \ne 2$ and $\eta \kappa ^{2} \nu ^{2} \ne 2/3$ for $\eta > 0$. Inserting $Z$ and $V$ in (\ref{simplyEOM4}) into the third equation in  (\ref{simplyEOM3})  we have a highly non-linear  differential equation, namely
\begin{eqnarray}
&& 2\varsigma \left( \frac{d^2Y}{dx^2} + 3 \frac{dY}{dx} + 4 k \eta  \chi \nu^2 e^{-2x} \right) + \frac{1}{Y}  \left( \frac{\chi}{\tilde{\chi }}  - \varsigma \frac{dY}{dx} \right) \left( \frac{dY}{dx} + 3Y - 2 k \eta \chi \nu^2 e^{-2x} +2\varepsilon \chi \nu ^{2} \right)^2  \nonumber\\
&& \quad  -  \frac{\chi}{\tilde{\chi }}  \left(  3 \frac{dY}{dx} + 9Y+   \frac{4 k \chi}{\kappa^2} \left(2 + \eta \kappa ^{2} \nu ^{2} \right)  e^{-2x} \right)  = 0 ~ , \label{HNLODE}
\end{eqnarray}
 which is quite complicated and difficult to solve analytically. Equation  (\ref{HNLODE}) is the master equation which  differs from that of  the scalar-tensor theory \cite{Kofinas:2015hla}. We also have 
\begin{eqnarray} 
\left(\frac{dx}{d\phi } \right)^{2} &=& Y(x)  ~  ,\nonumber\\
\left[\frac{d\ln \left(\rho f\right)}{dx} \right]^{2}  &=& \frac{Z}{Y}   ~  , \label{derivatphi}
\end{eqnarray} 
such that the metric (\ref{staticmet}) can be written as \cite{Kofinas:2015hla}
\begin{equation} \label{gengeom} 
ds^{2} =-f  dt^{2} +\frac{d\rho ^{2} }{\nu ^{2} \rho ^{2} Y} +\rho ^{2}  d\Omega ^{2}_k  ~ , 
\end{equation} 
which implies that both $Y$ and $Z$ must be positive definite. In the case at hand, the norm of the Riemann tensor (\ref{normRiem}) simplifies to
\begin{eqnarray} 
R^{\alpha\beta\mu\nu} R_{\alpha\beta\mu\nu} &=&  \frac{ \nu^2 \rho^2 Y}{ f^2 } \left( f_{\rho \rho} - \frac{f_{\rho}^2}{2f} + \frac{f_{\rho}}{2}  \left(\frac{2}{\rho}  + \frac{1}{Y}   \right) \right)^2 + \frac{2  f_{\rho}^2}{ f^2} \nu^4  \rho^2 Y^2 \nonumber\\
&&  + 2 \nu^4  \rho^2 Y^2  \left(\frac{2}{\rho}  + \frac{1}{Y}   \right)^2 + \frac{4}{\rho^4}  \left( k - \nu^2 \rho^2 Y \right)^2  ~  ,
\label{normRiem1} 
\end{eqnarray} 
where $f_{\rho} \equiv \frac{df}{d\rho}$ and  $ f_{\rho \rho} \equiv \frac{d^2 f}{d\rho^2}$. The Ricci scalar is simply given by
\begin{equation} \label{Ricciscalar} 
R = \left(-9 -  \sqrt{\frac{Z}{Y}} - \frac{2 \rho}{Y} \frac{dY}{d\rho} + \frac{2k}{\nu ^{2} \rho^2  Y} \right) \nu^2 Y ~ . 
\end{equation} 
%
\section{Solutions Near Boundaries}
\label{sec:solbond}
In this section we will show that for our model with the ansatz (\ref{simpelphi}), it is possible to have a hairy black hole solution. This can be seen in two regions, namely in the asymptotic region and the near-horizon region. As mentioned above, we have generally different situations than the case considered, for example in \cite{Gibbons:1996prl, Gunara:2010iu}. This is so because the scalar field are not frozen near the boundaries which implies the scalar potential cannot be extremized. 

\subsection{Around Asymptotic Region}
 In the asymptotic limit, namely at $r \to + \infty$, we may assume  $x \to + \infty$ such that at the zeroth order, the solution of (\ref{HNLODE}), i.e. $Y$ tends to a positive constant  $Y_0$, given by
\begin{equation} 
Y_0 = - \frac{2\varepsilon \nu^2 \chi \tilde{\chi}}{ 3 (  \tilde{\chi} -  \varsigma_1 \chi )} ~ ,
\end{equation} 
where $ \varsigma_1 = \pm 1$. By solving the second equation in (\ref{derivatphi}), the lapse function $f$  in (\ref{gengeom}) becomes
\begin{equation} 
f (\rho) =  A_0  \rho^{3\varsigma \varsigma_1 - 1} ~ ,
\end{equation} 
 with $A_0$ is a non-zero constant. In this limit, the effective geometry (\ref{gengeom})  becomes  a static spacetime of constant Ricci curvature with 
 \begin{equation} 
 R = - (9 + 3 \varsigma \varsigma_1) \nu^2 Y_0 ~ ,
 \end{equation} 
that can be split into
  \begin{equation} 
 R = \left\{ \begin{array}{ll}
   -12 \nu^2 Y_0 & \textrm{if $\varsigma \varsigma_1 =1$ (Einstein)}\\
    -6 \nu^2 Y_0 & \textrm{if $\varsigma \varsigma_1 = -1$ (non-Einstein)}
   \end{array} \right.~  .
   \end{equation} 
  The norm (\ref{normRiem1}) tends to be quadratic in $\rho$, namely
\begin{equation} 
R^{\alpha\beta\mu\nu} R_{\alpha\beta\mu\nu} = 2 \nu^4   Y_0^2 \rho ^{2} ~ ,  \label{normRiem2}
\end{equation} 
which shows that in this limit the spacetime is not maximally symmetric. We want to remark that in the case of $\varsigma \varsigma_1 =1$, the spacetime is not anti-de Sitter since the norm of Riemann tensor is not a constant, see for example in \cite{Joycebook}.\\
\indent The word ``effective'' means that the scalar field $\phi$ is not frozen in the region, i.e. at the lowest order the solution of the first equation in (\ref{derivatphi}) is given by
\begin{equation} 
\phi(\rho) = \phi_0 + Y_0^{-1/2} ~ {\mathrm{ln}}\rho ~ ,  \label{scalarinfty}
\end{equation} 
where $\phi_0$ is a constant. As a consequence the scalar potential $V(\phi)$ cannot be extremized, but it has a constant value at the zeroth order, given by
\begin{equation} 
V_0 =  -\frac{3 Y_0}{2 \chi}-\frac{\varepsilon \nu ^{2} }{2} ~ . 
\end{equation}
Now, we compute a mass-like quantity  of the black hole using the Komar integral defined in \cite{Gunara:2010iu, Kastor:2008cqg}. We obtain a non-constant mass-like quantity
\begin{equation} 
M(\rho, g_o,\tau) =  \left[ (3\varsigma \varsigma_1  + 4)  Y_0 ~\nu^2  \rho + C_1 \rho^{-3(\varsigma \varsigma_1  -1)/2}\right] \rho^2 A(g_o,\tau) ~ , \label{MassBH}
\end{equation}
%
%
where the constant $C_1$ is related to a non-trivial cohomology of ${\bf S}^2_k$ and assumed to be non-negative. This mass-like quantity satisfies partially  Bartnik's condition on quasi local mass \cite{Bartnik}, namely (i) Well-defined, (ii) $M > 0$, and (iii) Monotonicity. However, it diverges as $\rho \to \infty$ which is opposite to some quasi local mass formulae, for example, in \cite{Chrusciel}. Note that we cannot use some quasi local mass formula, like for example in \cite{Chrusciel}, since we do not have the flat space in the asymptotic region. The quantity $A(g_o,\tau)$ in (\ref{MassBH}) is the area of the surface  ${\bf S}^2_k$ given by \cite{Vanzo:1997gw}
\begin{equation} 
 A(g_o,\tau) =  4\pi |g_o -1| + |{\mathrm{Im}}\tau|  \delta (g_o,1)  ~ ,
\end{equation}
 with $g_o$ is the genus of the Riemann surface  ${\bf S}^2_k$ and $\tau$ is a complex number known as the Teichm\"uller complex parameter of the torus.\\
\indent  On the other hand, a scalar charge-like in this case is given by
\begin{equation} 
 Q(\rho) \equiv  - \frac{1}{2 A(g_o,\tau)} \int_{{\bf S}^2_k(\infty)} \frac{d\phi}{d\rho} \rho^2 dS = - \frac{\rho}{2} Y_0^{-1/2} ~ , \label{scalarcharge}
\end{equation}
%
 which has a negative value for large $\rho$. Similar to (\ref{MassBH}), it is called a scalar charge-like because it diverges as  $\rho \to \infty$.\\
 \indent We would like  to make some comments as follows. In the asymptotic region the Schwarzschild spacetime limit does not exist since the scalar does not converge to zero and moreover, the norm of Riemann tensor (\ref{normRiem2}) diverges. From (\ref{scalarinfty}) and (\ref{MassBH}), it can be easily seen that the mass-like quantity of the black hole determines the behavior of the scalar field where the mass-like is not a constant parameter but depends on the metric function $\rho$. For such a case we name the hair to be \textit{secondary-like}. Furthermore, the mass-like quantity in (\ref{MassBH})  can also be expressed in terms of the modulus of the scalar charge-like (\ref{scalarcharge}).


%

\subsection{Near Horizon}
 
 Now, let us consider the near-horizon geometry. First,  we write down the mean curvature $H$ of  ${\bf S}^2_k \subseteq \Sigma^3$ 
  \begin{equation} 
 H = \nu  Y^{1/2} ~  ,
 \end{equation} 
  where $\Sigma^3$ is the hypersurface at $t = \text{constant}$. On the horizon,  the surfaces ${\bf S}^2_k$ have to be minimal for all $k$, namely $H = 0$, and therefore $Y = 0$. In order to evade the blow up of the metric (\ref{gengeom}) and the ansatz (\ref{simpelphi}), around the region the function $\rho(r)$ should have the form
  \begin{equation} 
  \rho(r) = \rho(r_h) + \epsilon^{\frac{1}{2}} (r - r_h) ~  , \label{BHcon}
  \end{equation} 
 where $r_h$ is the radius of the horizon, $\rho(r_h) > 0$,  and $Y(r \to r_h) = \epsilon$, $0 < \epsilon \ll 1$. From (\ref{simplyEOM4}), we find 
 \begin{equation} 
   \rho(r_h) = \left(\varepsilon k \eta\right)^{1/2} ~  ,
  \end{equation}
 therefore it forbids the 2-torus (or $k =0$). The two possible cases are $k = \varepsilon$ and $\eta > 0$ or  $k = - \varepsilon$ and $\eta < 0$. In this limit the lapse function $f(r)$ becomes a constant and the near-horizon geometry has to be $\lR^{1,1} \times {\bf S}^2_k$ where $\lR^{1,1}$ is the 2-Minkowskian surface and $k = \pm 1$ \cite{Kunduri:2007} whose scalar curvature is given by
  \begin{equation} 
R = \frac{2 \varepsilon}{\eta} ~  . \label{scalarcurvhor}
  \end{equation}
 From (\ref{derivatphi}), near this region the scalar field has the form
 \begin{equation} 
  \phi(r) = \phi_h + \frac{1}{\rho(r_h)}  (r -r_h)   ~  , \label{scalarfuncnearhor}
  \end{equation}
with $\phi_h$ is a constant. From (\ref{derivatphi}) and (\ref{BHcon}), we find that the surface gravity 
 \begin{equation} 
  \kappa_s =    \frac{f'}{2f^{1/2}} \nu \rho Y^{1/2} |_{r=r_h} = 0 ~  , \label{surfacegrav}
  \end{equation}
on the region which follows that we have an ultra-cold black hole with temperature $T=0 ~ {\mathrm K}$ \cite{Kallosh:1992} in the case at hand.  The scalar potential $V$ in (\ref{simplyEOM4}) becomes constant, namely
 \begin{equation} 
V_h = - \frac{\varepsilon \nu^2}{\eta \kappa^2} \left( 1+ \eta \kappa^2 \nu^2\right) ~  .
 \end{equation}
Since the scalar potential $V$ is not extremized, we then have
\begin{equation} 
R \ne - 4 V_h  ~  .
 \end{equation}
\\
\indent We close this section by making some remarks as follows. First, if the order of $\varepsilon$ in (\ref{BHcon}) were $p$ with $p \ne \frac{1}{2}$, then this would imply $\nu = \varepsilon^{p-\frac{1}{2}}$. For $p < \frac{1}{2}$, the scalar field $\phi$ and $V_h$ would diverge, while the scalar field $\phi$  would be frozen and $V_h \to 0$ for $p > \frac{1}{2}$. The latter case turns out to be forbidden since it leads to the inconsistency (\ref{scalarcurvhor}) and the condition $R = - 4 V_h $. Second, it is also possible that the metric (\ref{gengeom}) does not have any minimal surface, \textit{i.e.} $H \ne 0$. If this is the case then the spacetime could be smooth everywhere or it might have a naked singularity at the origin. \\
\indent Comparing (\ref{scalarinfty}) and (\ref{scalarfuncnearhor}) we can conclude that it is impossible to match up the scalar in both region up to perturbative expansion if the setup (\ref{simpelphi}) is fulfilled. However, we could also take the setup discussed in  \cite{Gibbons:1996prl, Gunara:2010iu} where in the asymptotic region the scalar has the form
\begin{equation} 
 \phi(\rho)= \phi_{\infty} + Q/ \rho  ~ ,
 \end{equation}
 while at the horizon the scalar becomes constant, namely, $\phi = \phi_h$ with constant scalar charge $Q$. In the latter case, the scalar potential $V(\phi)$ in (\ref{ActionNMDCV}) should be extremized at the boundaries and moreover, we could have $\phi_{\infty} = \phi_h$. Thus, we have a critical black hole where the asymptotic geometry might be flat Minkowskian or Einstein depending on the value of $V(\phi_{\infty})$. This latter case is a complement to our solution considered in this paper since it does not satisfy (\ref{simpelphi}).
    
%
%


\section{Conclusions}
\label{sec:conc}
So far we have considered a family of static black holes in four dimensional Einstein-Klein-Gordon theory with non minimal derivative coupling  and the scalar potential is turned on. At the beginning, the black holes  have unusual topology, namely, the two-surfaces are maximally symmetric with $k = 0, \pm 1$, see (\ref{staticmet}).  In particular, we take a simple model satisfying the ansatz (\ref{simpelphi}) which simplifies the Einstein field equation (\ref{Einsteineqstatic})  and the scalar field equation of motions (\ref{scalareqomstatic}) which are coupled and highly nonlinear. This setup further implies that these coupled differential equations reduce into a single highly nonlinear differential equation, namely equation (\ref{HNLODE}).\\
\indent First, we showed that in the asymptotic limit, the metric function $Y$ introduced in (\ref{derivatphi}) tends to constant $Y_0$ but the scalar $\phi$ turns out to be not fixed. tAs a consequence of his latter case, the scalar potential $V$ in (\ref{ActionNMDCV}) is not extremized. The effective geometry is the space of negative constant scalar curvature. The mass-like quantity depends on $\rho$ given in (\ref{MassBH})  showing that the hair to be secondary-like since the mass-like quantity can be expressed  in terms of scalar charge-like.

Then, near-horizon, in order to have a regular metric the metric function $\rho(r)$ should take the form (\ref{BHcon}). The near-horizon geometry has the form $\lR^{1,1} \times {\bf S}^2_k$ where $\lR^{1,1}$ is the 2-Minkowskian surface and $k = \pm 1$. In other words, the torus with $k=0$ is excluded. The scalar field in this case has a linear form with respect to the radial coordinates $r$ which implies that the surface gravity vanishes in this region. So, our black hole is an ultra-cold type. Again, the scalar potential $V$ cannot be extremized. 

\appendix

\section{The Next Leading Order}

In this section, we  consider the first order solutions in the asymptotic limit using a perturbative ansatz 
%
 \begin{equation} 
 Y = Y_0 +   Y_1 ~  ,
 \end{equation} 
with $|Y_0|   \gg |Y_1(x)| $ showing that we have a regular solution up to the perturbative correction.   Such a setup  simplifies the master equation (\ref{HNLODE}) into the following linear equation
 \begin{eqnarray} 
&& 2 \tilde{\chi} \frac{d^2Y_1}{dx^2} + 3 \left(4 \tilde{\chi} -  \chi -\frac{2 \nu^2 \chi^3}{\chi -  \tilde{\chi}} \right) \frac{dY_1}{dx} -18 \left( \chi -  \tilde{\chi} \right) Y_1 \nonumber\\
&& + 4 \tilde{\chi} \left[ \left(2 + \eta \kappa^2 \nu^2 \right) \frac{\chi^2 k}{\tilde{\chi} \kappa^2 } -\eta \nu^2 \chi k  \right] e^{-2x} + 4 \tilde{\chi} \chi^2 \eta^2 \nu^4 k^2 e^{-4x} = 0 ~  , \label{linearHNLODE}
 \end{eqnarray} 
where we  have simply set $\varsigma = \varsigma_1 = 1$. Moreover, we also take $e^{-mx} Y_1(x) \to 0$ and $e^{-mx} dY_1/dx \to 0$ for $m \ge 2$ around $ x \to +\infty$. A special class of solution of (\ref{linearHNLODE}) has the form
 \begin{eqnarray} 
Y_1 &=& C_1 S[\lambda_1] ~ e^{\lambda_1 x} +  D_1 ~ e^{\lambda_2 x} - \frac{2 \left( \left(2 + \eta \kappa^2 \nu^2 \right) \frac{\chi^2 k}{\tilde{\chi} \kappa^2 } -\eta \nu^2 \chi k  \right)}{(\lambda_1 +2) (\lambda_2 +2)}  e^{-2x} \nonumber\\
&& - \frac{2 \chi^2 \eta^2 \nu^4 k^2 }{(\lambda_1 +4) (\lambda_2 + 4)}  e^{-4x} ~  , \label{sollinearHNLODE}
 \end{eqnarray} 
where $C_1, D_1 \in \lR$ with
 \begin{equation} 
 \lambda_{1, 2} =  \frac{3}{4} \left( \frac{\chi}{\tilde{\chi}} - 4  + \frac{2 \varepsilon \nu^2 \chi^3}{\tilde{\chi} ( \chi -  \tilde{\chi}) } \pm \left[ \left(  \frac{\chi}{\tilde{\chi}} - 4  + \frac{2 \varepsilon \nu^2 \chi^3}{ \tilde{\chi} ( \chi -  \tilde{\chi}) } \right)^2 + 16 \left(  \frac{\chi}{\tilde{\chi}} - 1 \right)   \right]^{1/2}\right) ~  , \label{eigenwert}
 \end{equation} 
and
\begin{equation} 
S[\lambda_1] = \left\{ \begin{array}{ll}
 0 & \textrm{if $\lambda_1 > 0$}\\
 1 & \textrm{if $\lambda_1 < 0$}
 \end{array} \right.~  ,
 \end{equation} 
The values of $\lambda_1$ or $\lambda_2$ should be negative, since $\lim_{ x \to +\infty} Y_1(x) \to 0$, and $ -2 < \lambda_{1, 2} < 0$. The solution (\ref{sollinearHNLODE}) belongs to the following two cases. First, by simply taking $0 < \chi /\tilde{\chi} < 1$ and $\eta > 0$, the analysis on (\ref{eigenwert}) results $\varepsilon = 1$ and
\begin{equation} 
2 < \eta \kappa^2 \nu^2 < \frac{2}{3} (1 + 4\kappa^2) ~ . \label{intervalsol1}
 \end{equation}
In this case,  both  $\lambda_1$ and $\lambda_2$ are negative and we exclude the large $\kappa$ limit case.  Second, we simply take $ \chi /\tilde{\chi}  > 1$ and $\eta < 0$. Analyzing (\ref{eigenwert}), we find  $\varepsilon = 1$ and
\begin{equation} 
 2 - \frac{32}{9}\kappa^2  < \eta \kappa^2 \nu^2 < 0 ~ , \label{intervalsol2}
 \end{equation} 
 and only  $\lambda_2$ is negative.\\
 \indent Solving the second equation in (\ref{derivatphi}), we obtain the first order lapse function $f$ 
 \begin{equation} 
 f (\rho) =  \hat{A}_0 Y_0^{\varsigma \tilde{\chi}/\chi}  \rho^{3\varsigma \varsigma_1 - 1 } \left( 1 + \frac{\varsigma \tilde{\chi} Y_1}{\chi Y_0}  \right) e^{k\varsigma \eta  \tilde{\chi} \nu^2 / Y_0 \rho^2}  ~ , \label{lapsefuncasymp}
 \end{equation} 
 while the first equation gives us  
 \begin{eqnarray} 
  \phi (\rho) &=&  \phi_0 + \int \frac{dx}{\sqrt{Y_0 + Y_1}} \nonumber\\
  &\approx&  \phi_0 + Y_0^{-1/2} ~ {\mathrm{ln}}\rho -\frac{1}{2} Y_0^{-3/2} \Bigg( \frac{C_1}{\lambda_1} S[\lambda_1] ~ \rho^{\lambda_1} +  \frac{D_1}{\lambda_2} ~ \rho^{\lambda_2}\nonumber\\
  &&  + \frac{ \left( \left(2 + \eta \kappa^2 \nu^2 \right) \frac{\chi^2 k}{\tilde{\chi} \kappa^2 } -\eta \nu^2 \chi k  \right)}{(\lambda_1 +2) (\lambda_2 +2)}  \rho^{-2} + \frac{ \chi^2 \eta^2 \nu^4 k^2 }{2(\lambda_1 +4) (\lambda_2 + 4)}  \rho^{-4}   \Bigg) ~ . \label{scalarfuncasymp}
  \end{eqnarray} 
Equations (\ref{lapsefuncasymp}) and (\ref{scalarfuncasymp}) show that we have a different set of solutions compared to \cite{Sotiriou:2014pfa}. From the second equation in (\ref{simplyEOM4}), we get the first order scalar potential
\begin{eqnarray} 
V &=&  -\frac{3 Y_0}{2 \chi}-\frac{\varepsilon \nu ^{2} }{2} -\frac{1}{2\chi } \left( C_1 (\lambda_1 +3) S[\lambda_1] ~ \rho^{\lambda_1} +  D_1 (\lambda_2 +3) ~ \rho^{\lambda_2}  \right) \nonumber\\
&& + \left(\frac{  \left(2 + \eta \kappa^2 \nu^2 \right) \frac{\chi^2 k}{\tilde{\chi} \kappa^2 } -\eta \nu^2 \chi k  }{ \chi (\lambda_1 +2) (\lambda_2 +2)} -\frac{k}{2 \kappa ^{2} } (2 + \eta \kappa ^{2} \nu ^{2})  \right)   \rho^{-2} + \frac{ \chi^2 \eta^2 \nu^4 k^2 }{ \chi (\lambda_1 +4) (\lambda_2 - 4)}  \rho^{-4} ~ .\nonumber\\
\end{eqnarray}

%


\section*{Acknowledgments}

We would like to thank M. Satriawan for correcting grammar and anonymous referee for suggestion of the paper. The work in this paper is supported by Riset KK ITB 2015-2017 and PDUPT Kemenristekdikti-ITB 2015-2018.




\end{document}